\documentclass[11pt]{article}
\pdfoutput=1
\usepackage[centertags]{amsmath}
\usepackage[square, comma, sort&compress,numbers]{natbib}
\usepackage{array,multirow}
\numberwithin{equation}{section}
\usepackage{amssymb}
\usepackage{graphicx}
\usepackage{color}

\usepackage{epsfig}
\usepackage{bbold}

\usepackage{wrapfig}
\usepackage{float}
\usepackage{soul}

\usepackage{tkz-euclide}

\usepackage{tikz,pgf}
\usetikzlibrary{shapes}
\usetikzlibrary{calc}
\usetikzlibrary{decorations.pathmorphing}
\usetikzlibrary{decorations.pathreplacing,shapes.misc}
\usetikzlibrary{positioning}
\usetikzlibrary{arrows}
\usetikzlibrary{decorations.markings}
\usetikzlibrary{shadings}

\usetikzlibrary{intersections}

\def\be{\begin{equation}}
\def\ee{\end{equation}}
\def\ba{\begin{array}}
\def\ea{\end{array}}

\def\dps{\displaystyle}
\newcommand{\half}{\frac{1}{2}}

\def\1{\tilde{1}}
\def\2{\tilde{2}}
\def\3{\tilde{3}}

\newdimen\tableauside\tableauside=1.0ex
\newdimen\tableaurule\tableaurule=0.4pt
\newdimen\tableaustep
\def\phantomhrule#1{\hbox{\vbox to0pt{\hrule height\tableaurule
width#1\vss}}}
\def\phantomvrule#1{\vbox{\hbox to0pt{\vrule width\tableaurule
height#1\hss}}}
\def\sqr{\vbox{%
  \phantomhrule\tableaustep

\hbox{\phantomvrule\tableaustep\kern\tableaustep\phantomvrule\tableaustep}%
  \hbox{\vbox{\phantomhrule\tableauside}\kern-\tableaurule}}}
\def\squares#1{\hbox{\count0=#1\noindent\loop\sqr
  \advance\count0 by-1 \ifnum\count0>0\repeat}}
\def\tableau#1{\vcenter{\offinterlineskip
  \tableaustep=\tableauside\advance\tableaustep by-\tableaurule
  \kern\normallineskip\hbox
    {\kern\normallineskip\vbox
      {\gettableau#1 0 }%
     \kern\normallineskip\kern\tableaurule}%
  \kern\normallineskip\kern\tableaurule}}
\def\gettableau#1 {\ifnum#1=0\let\next=\null\else
  \squares{#1}\let\next=\gettableau\fi\next}

\newcommand{\bref}[1]{\textbf{\ref{#1}}}

\def\cO{\mathcal{O}}

\numberwithin{equation}{section} \makeatletter
\@addtoreset{equation}{section}

\def\be{\begin{equation}}
\def\ee{\end{equation}}
\def\ba{\begin{array}}
\def\ea{\end{array}}

\def\dps{\displaystyle}

\def\ba{\begin{array}}
\def\ea{\end{array}}

\def\dps{\displaystyle}

\usepackage{jheppub}
\makeatletter
\def\@fpheader{\vspace{-.1cm}}
\makeatother

\title{ Large-$c$ conformal $(n \leq 6)$-point blocks with superlight weights and holographic Steiner trees }
\author[a]{Mikhail\ Pavlov}

\affiliation[a]{I.E. Tamm Department of Theoretical Physics, \\P.N. Lebedev Physical
Institute,\\ Leninsky ave. 53, 119991 Moscow, Russia}
\emailAdd{pavlov@lpi.ru}

\abstract{ In this note we study CFT$_2$ Virasoro conformal blocks with heavy operators in the  large-$c$ limit in the context of AdS$_3$/CFT$_2$ correspondence. We compute the lengths of the holographic Steiner trees dual to the $5$-point and $6$-point conformal blocks using the superlight approximation when one or more dimensions are much less than the others. These results are generalized for $N$-point holographic Steiner trees dual to $(N+1)$-point conformal blocks with superlight weights. }

\makeatletter
\def\@fpheader{\vspace{-.1cm}}
\makeatother

\begin{document}

\maketitle
\flushbottom

\section{Introduction}

The study of the AdS/CFT correspondence \cite{Maldacena:1997re, Aharony:1999ti} provides many new ideas and fruitful observations related to computations in QFT. In the case of AdS$_3$/CFT$_2$ correspondence it is essential to consider the large-$c$ limit which corresponds to the weak gravitational coupling in the bulk according to Brown-Henneaux formula \cite{Brown:1986nw}. One of the most elaborated issues is the correspondence between Virasoro conformal blocks with heavy operators in the large-$c$ limit and probe particles propagating in the AdS$_3$ background with conical defects originally obtained for lower-point blocks \cite{Hartman:2013mia, Fitzpatrick:2014vua,  Hijano:2015rla, Fitzpatrick:2015zha, Alkalaev:2015wia, Hijano:2015qja, Alkalaev:2015lca}.\footnote{Other recent related research focuses on $p$-adic AdS/CFT correspondence \cite{Parikh:2019ygo, Jepsen:2019svc,Jepsen:2020kfd}, entanglement entropy \cite{Calabrese:2004eu, Calabrese:2009qy} and OTOC computations \cite{Anous:2016kss,Kusuki:2019gjs, Anous:2020vtw}. } The large-$c$  $n$-point conformal blocks were studied in \cite{Banerjee:2016qca, Chen:2016dfb, Alkalaev:2016rjl, Kusuki:2019gjs}. However, exact expressions for large-$c$ conformal blocks are still unknown.

In this work, we continue to study large-$c$ conformal blocks as holographic Steiner trees on the Poincare disk \cite{Alkalaev:2018nik}.
We consider holographic Steiner trees with $N=4$ and $N=5$ endpoints in the superlight approximation where one or more weights are much less than the others. Their lengths are calculated by making use of the hyperbolic trigonometry relations. On the boundary, such Steiner trees are dual to the large-$c$ conformal blocks with superlight operators \cite{Alkalaev:2015wia}.  Also, we find the lengths of $(2M+1)$ holographic Steiner trees in the superlight approximation corresponding to the $(2M+2)$-point large-$c$ conformal blocks with superlight operators.

The paper is organized as follows. In Section \bref{sec:HT} we study the  Steiner tree problem on the Poincare disk and calculate the lengths of the holographic Steiner trees with $N= 4$ and $N=5$ endpoints in the superlight approximation. Section \bref{sec:semi-CFT} applies the monodromy method to calculate the large-$c$ conformal blocks in the heavy-light approximation extended further by the superlight approximation. Here, we show the holographic correspondence relation between large-$c$ conformal blocks and the lengths of the Steiner trees obtained in Section \bref{subsec:ex}. Concluding Section \bref{sec:con} summarizes our results.

\section{Holographic Steiner trees on the Poincare disk}
 \label{sec:HT}
 In the context of AdS$_3$/CFT$_2$ correspondence the Poincare disk with an angle deficit arises as a constant-time slice of AdS$_3$ space \cite{ Alkalaev:2015wia, Hijano:2015rla}. In this section we focus on the Steiner tree problem on the Poincare disk\footnote{The Euclidean Steiner tree problem in context of QFT is considered in \cite{Avdoshkin:2018lfs}.} for the special class of trees called holographic \cite{Alkalaev:2018nik}. We use hyperbolic trigonometry to calculate particular holographic Steiner trees with $N = 3, 4, 5$ endpoints and then generalize these results to $N$-point Steiner trees.

\subsection{The Steiner problem on the Poincare disk}
\label{sec:st}

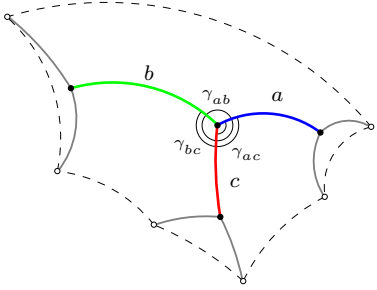
\begin{figure}[H]
\centering
\begin{tikzpicture}[line width=0.3pt,scale=1.90]

  \tkzDefPoint(0.8,-0.6){z_{3}}
  \tkzDefPoint(-0.945,-0.29){z_{1}}
  \tkzDefPoint(0.1,-1.191){z_{2}}

  \tkzDefPoint(0.08,-0.55){x}

  \tkzDefPoint(-1.39,0.21){gv1}
  \tkzDefPoint(-1.04,-0.87){gv2}

  \tkzDefPoint(-0.365,-1.245){gv3}
  \tkzDefPoint(0.26,-1.64){gv4}

  \tkzDefPoint(0.83,-1.05){gv5}
  \tkzDefPoint(1.155,-0.56){gv6}

  \tkzDefPoint (-0.655, -1.31){c_c_1}
  \tkzDefPoint (2.49, -0.79){c_c_2}
  \tkzDefPoint (0.4,-1.13){c_c_3}

  \tkzDefPoint (-2.6, -1.3){c_g_11}
  \tkzDefPoint (-1.5, -0.5){c_g_12}

  \tkzDefPoint (0, -2.3){c_g_21}
  \tkzDefPoint (-1.5, -2){c_g_22}

  \tkzDefPoint (1.2, -0.8){c_g_31}
  \tkzDefPoint (1, -0.8){c_g_32}

   \tkzDefPoint (0.08,-0.61){a1}
   \tkzDefPoint (-0.03,-0.45){a2}
   \tkzDefPoint (0.04,-0.45){a3}

 \tkzDrawArc[rotate,color=black,line width  = 0.1](x,a1)(120)
  \tkzDrawArc[rotate,color=black,line width  = 0.1](x,a2)(245)
   \tkzDrawArc[rotate,color=black,line width  = 0.1](x,a3)(360)

 \tkzDrawArc[rotate,color=green,line width  = 1.0](c_c_1,z_{1})(-60)
 \tkzDrawArc[rotate,color=red, line width  = 1.0](c_c_2,z_{2})(-15)
 \tkzDrawArc[rotate,color=blue,line width  = 1.0](c_c_3,z_{3})(66)

 \tkzDrawArc[rotate,color=gray, line width  = 0.7](c_g_11,z_{1})(20)
 \tkzDrawArc[rotate,color=gray, line width  = 0.7](c_g_12,z_{1})(-60)

 \tkzDrawArc[rotate,color=gray, line width  = 0.7](c_g_21,z_{2})(25)
 \tkzDrawArc[rotate,color=gray, line width  = 0.7](c_g_22,z_{2})(-15)

 \tkzDrawArc[rotate,color=gray, line width  = 0.7](c_g_31,z_{3})(60)
 \tkzDrawArc[rotate,color=gray, line width  = 0.7](c_g_32,z_{3})(-80)


 \tkzDefPoint (-0.7, -2.1){f1}
  \tkzDefPoint(1.26, -0.98){f2}
  \tkzDefPoint(1.43, -2.2){f3}
  \tkzDefPoint(-1.16,-3.2){f4}
  \tkzDefPoint(-1.16,-1.88){f5}
  \tkzDefPoint(-2.16,-0.64){f6}

 \tkzDrawArc[rotate,dashed, color=black,line width  = 0.3](f1,gv1)(-67)
  \tkzDrawArc[rotate,dashed, color=black,line width  = 0.3](f2,gv6)(88)
  \tkzDrawArc[rotate,dashed, color=black,line width  = 0.3](f3,gv5)(38)
  \tkzDrawArc[rotate,dashed, color=black,line width  = 0.3](f4,gv4)(20)
  \tkzDrawArc[rotate,dashed, color=black,line width  = 0.3](f5,gv3)(45)
  \tkzDrawArc[rotate,dashed, color=black,line width  = 0.3](f6,gv2)(59)

 \tkzDrawPoints[color=black,fill=white,size=2](gv1,gv2,gv3,gv4,gv5,gv6)
\tkzDrawPoints[color=black,fill=black,size=2](z_{1},z_{2},z_{3}, x)

 \draw (0.2, -0.95) node {\scriptsize$c$};
  \draw (0.5, -0.35) node {\scriptsize $a$};
  \draw (-0.4, -0.18) node {\scriptsize$b$};

  \draw (0.29, -0.75) node {\tiny $\gamma_{_{ac}}$};
   \draw (-0.12, -0.7) node {\tiny $\gamma_{_{bc}}$};
    \draw (0.08, -0.35) node {\tiny $\gamma_{_{ab}}$};

\end{tikzpicture}
\vspace{-3cm}
\caption{ $N=6$ Steiner tree. FT points are indicated by black points, different colors correspond to different weights, the angles are given by formula \eqref{cooos}. The outer hexagon is shown in dashed lines.}
\label{vertex}
\end{figure}

\paragraph{The Poincare disk.} Let $\mathbb{D}_{\alpha}$ denote the Poincare disk with the angle deficit which is parametrized by $\alpha \in (0, 1]$. In complex coordinates $(z, \bar{z})$ it is defined as $\mathbb{D}_{\alpha} = \{|z|<1, \text{arg} (z)  \in[0, 2\pi \alpha)\}$ and the boundary is a part of the circle $\partial \mathbb{D}_{\alpha} = \{|z|=1, \text{arg}(z) \in[0, 2\pi \alpha)\}$. After reparameterization $\text{arg} (z)  \rightarrow \alpha \;  \text{arg} (z) $ we obtain the Poincare disk model $\mathbb{D}$. In what follows we do all calculations on the Poincare disk and then recover parameter $\alpha$.
The length of a geodesic segment between two points $z_1$ and $z_2$ is given by
\be
\label{L}
L_{\mathbb{D}}(z_1, z_2) = \log \frac{1+u}{1-u}, \qquad u = \frac{|z_1 - z_2|}{|1- \bar z_1 z_2|}\;.
\ee
The regularized length (see e.g. appendix A in \cite{Alkalaev:2018nik} for details) of the geodesic connecting two boundary endpoints $z_i = \exp[i w_i] \text{ and } z_j = \exp[i w_j]$ takes the form
 \be\label{r1}
  L_{\mathbb{D}}^{\varepsilon} (w_i, w_j ) = a_{ij} -2 \log \varepsilon  \;, \qquad a_{ij} \equiv \log \left[4 \sin^2 w_{ij}\right]  \;, \qquad w_{ij} \equiv \frac{w_i- w_j}{2}\;,
\ee where the regulator $ \varepsilon \rightarrow 0+ $. The regularized length of the geodesic connecting the bulk point $z = r \exp[ i \varphi]$ and the boundary point $z_i = \exp[ i w_i]$ is given by
  \be\label{r2}
 L^{\varepsilon}_{\mathbb{D}}(w_i, r, \varphi) = b - \log \varepsilon \;, \qquad b \equiv  \log \frac{ 2 \left(r^2-2 r \cos(\varphi-w_i )+1\right)}{1-r^2}  \;.
 \ee We denote by $ L_{\mathbb{D}} $ the finite part of the regularized length on the Poincare disk which is obtained by discarding the $\varepsilon$-dependent terms in \eqref{r1} and \eqref{r2}.
 \paragraph{Steiner trees.} Given $N$ points (outer vertices) belonging to $\mathbb{D}$ or $\partial \mathbb{D}$ we consider a connected tree $G_N$ with $N$ outer edges attached to outer vertices and $N-3$ inner edges. The outer and inner edges are connected to each other at $N-2$ trivalent inner vertices. The weighted length of $G_N$ reads
\be
\label{minimum}
 L^N_{\mathbb{D}} = \sum_{\{\text{outer edges}\}} \epsilon_i L_i + \sum_{ \{\text{inner edges}\}} \tilde \epsilon_j \tilde L_j \;,
\ee where $\epsilon_i$ and $\tilde \epsilon_j$ are weights of outer and inner edges, respectively. The Steiner problem is to find positions of inner vertices for given tree and weights such that the weighted length \eqref{minimum} is minimal.\footnote{For more detailed analysis, see \cite{Ivanov_Tuzhilin, Zachos, Anderson}.}  In this case the inner vertices are called Fermat–Torricelli (FT) points and $G_N$ is called the Steiner tree (see Fig. \bref{vertex}).
Also, for further purposes one can consider a hyperbolic  $N$-gon with corners at the outer vertices of the Steiner tree (outer polygon).
One can show that the angles between edges with weights $\epsilon_a, \epsilon_b, \epsilon_c$ intersecting at FT point  are given by
\begin{equation}\label{cooos}
       \cos \gamma_{ac} = \frac{-\epsilon^2_c - \epsilon^2_b + \epsilon^2_a}{2 \epsilon_a \epsilon_c}\;,\quad
       \cos \gamma_{bc} = \frac{-\epsilon^2_c + \epsilon^2_b - \epsilon^2_a}{2 \epsilon_c \epsilon_b}\;,\quad
        \cos \gamma_{ab} = \frac{\epsilon^2_c - \epsilon^2_b - \epsilon^2_a}{2 \epsilon_a \epsilon_b} \;,
\end{equation} supplemented by the triangle inequalities
\be
\label{triangle}
\epsilon_a + \epsilon_b \geq \epsilon_c\;,
\qquad
\epsilon_a + \epsilon_c \geq \epsilon_b\;,
\qquad
\epsilon_b + \epsilon_c \geq \epsilon_a\;.
\ee
The relations \eqref{cooos} and \eqref{triangle} follow from the requirement that the Steiner tree has a minimal length and fix the positions of the FT points.

In what follows we focus on two types of Steiner trees \cite{Alkalaev:2018nik}: 1) $N$ boundary endpoints, 2) $N-1$ boundary endpoints and one endpoint in the center of $\mathbb{D}$. We will refer to them as ideal and non-ideal holographic Steiner trees,\footnote{These trees are characterized by a certain topology and on the boundary side turn out to be dual to the s-channel classical conformal blocks. Other block/tree topologies were studied in \cite{Chen:2016dfb, Banerjee:2016qca}. } respectively.

\paragraph{Superlight approximation.}
\label{LWA} Suppose  now that one of the three weights in \eqref{cooos} is much less than the other two, which are assumed to be equal,
\be
\label{sl}
\epsilon_c \ll \epsilon_a = \epsilon_b : \qquad  ~~~~  \gamma_{ab} = \pi\;, \qquad  \gamma_{ac} = \gamma_{bc} = \pi/2 \;.
\ee
We see that two edges of the vertex merge into a single geodesic segment while the third edge stretches in a perpendicular direction. Then, the case of three arbitrary weights can be regarded as a perturbation of this configuration in the small parameter $\epsilon_c$.

\paragraph{Hyperbolic trigonometry.}
\label{subsec:triangles}

The lengths of the edges of the Steiner tree in \eqref{minimum} are determined by the coordinates of the FT points. For $N=3$ Steiner trees the coordinates of the FT point can be calculated explicitly but for  case $N\geq4$ the analysis becomes much more complicated. However, the lengths of the edges of  Steiner trees can be found using hyperbolic trigonometry. For example, $N=3$ Steiner trees cut the outer triangle into three triangles and the edges of the trees can be considered as the sides of the hyperbolic triangles. Here we provide the hyperbolic trigonometry relations that will be useful in calculating the edge lengths of holographic Steiner trees.

 Given a hyperbolic triangle with sides $A, B, C$ and interior angles $\alpha, \beta, \gamma$ opposite to $A, B, C$ the first and second cosine theorem, and the sine theorem read as \vspace{-0.5mm}
\begin{equation}\label{cos_th}
\ba{c}
\vspace{-4mm}
 \dps \cosh A = \cosh B  \cosh C - \sinh B \sinh C \cos \alpha\;, \\
  \\ \vspace{-3mm}
 \dps   \cosh C \sin \alpha \sin \beta = \cos \gamma + \cos \alpha \cos \beta \;, \;\;\;\;\;\;\;\;\;\;\; \\
\\
  \dps \frac{\sinh A}{\sin \alpha } = \frac{\sinh B}{\sin \beta} =  \frac{\sinh C}{\sin \gamma}\;. \;\;\;\;\;\;\;\;\;\;\;\;\;\;\;\;\;\;\;\;\;\;\;\;\;\;\;\;
\ea
\end{equation}When one of the vertices is on the boundary ($\beta = 0$), the first cosine law can be cast into the form
\be\label{cos_th1}
  \exp[A] =  \exp[C] (\cosh B  - \sinh B \cos \alpha) + \cO(\varepsilon)\;,
\ee where $A$ and $C$ denote the regularized lengths of sides connected to the vertex. For two vertices on the boundary ($\beta = \gamma = 0$) the regularized lengths $A, B, C$ are related as
\begin{equation}
\label{cos_th2}
   A = B +  C + 2 \log \sin \frac{\alpha}{2}+ \cO(\varepsilon)\;.
   \end{equation}

\subsection{Examples of Steiner trees}
\label{subsec:ex}
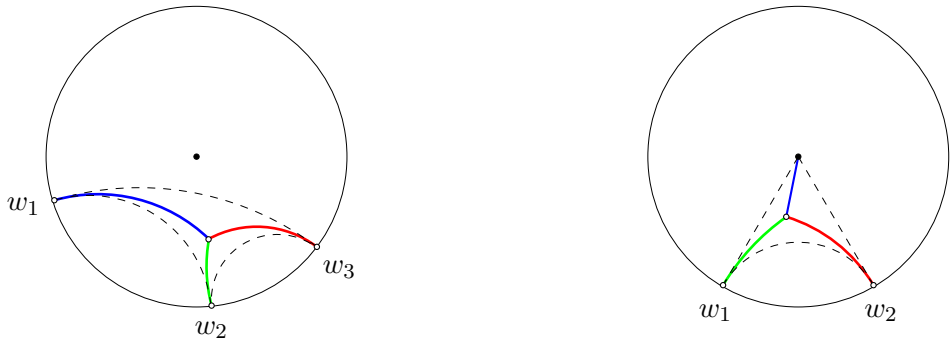
\begin{figure}[H]
\centering
\begin{tikzpicture}[scale=2.]
 \tkzDefPoint(6,0){OO}
  \tkzDefPoint(7,0){B}
  \tkzDrawCircle[line width  = 0.3](OO,B)

  \tkzDefPoint(2,0){OOO}
  \tkzDefPoint(3,0){C}
  \tkzDrawCircle[line width  = 0.3](OOO,C)

  \tkzDefPoint(6.5,-0.855){w1}
  \tkzDefPoint(5.5,-0.855){01}
  \tkzDefPoint(5.92,-0.4){x1}

  \tkzDefPoint(6,-1.158){a}

 \tkzDefPoint(6.777,-1.615){pp_1}
 \tkzDefPoint(5.618,-1.382){pp_2}


 \tkzDrawArc[rotate,color=red,line width  = 1.0](pp_2,w1)(42)
  \tkzDrawArc[rotate,color=green,line width  = 1.0](pp_1,01)(-24)
  \draw[blue,thick](6,0) -- (5.92,-0.4);

\tkzDefPoint(6,-1.15){dr}

\tkzDrawArc[dashed, rotate,color=black,line width  = 0.3](dr,w1)(122)
\draw[dashed,color = black, line width  = 0.3] (6, 0) -- (6.5,-0.855);
\draw[dashed,color = black, line width  = 0.3] (6, 0) -- (5.5,-0.855);

 \tkzDefPoint(2.8,-0.6){z_{3}}
  \tkzDefPoint(1.055,-0.29){z_{1}}
  \tkzDefPoint(2.1,-0.991){z_{2}}
  \tkzDefPoint (1.345, -1.31){c_c_1}
  \tkzDefPoint (3.1, -0.73){c_c_2}
  \tkzDefPoint (2.4,-1.13){c_c_3}

 \tkzDefPoint (1.283, -1.075){c_o_1}
  \tkzDefPoint (2.537, -0.956){c_o_2}
  \tkzDefPoint (1.62,-2.17){c_o_3}

  \tkzDefPoint(2.08,-0.55){xx1}

  \tkzDrawArc[rotate,color=blue,line width  = 1.0](c_c_1,z_{1})(-60)
  \tkzDrawArc[rotate,color=green,line width  = 1.0](c_c_2,z_{2})(-25)
  \tkzDrawArc[rotate,color=red,line width  = 1.0](c_c_3,z_{3})(66)

\tkzDrawArc[rotate,dashed, color=black,line width  = 0.3](c_o_1,z_{1})(-100)
  \tkzDrawArc[rotate,dashed, color=black,line width  = 0.3](c_o_2,z_{2})(-125)
  \tkzDrawArc[rotate,dashed, color=black,line width  = 0.3](c_o_3,z_{3})(50)

 \tkzDrawPoints[color=black,fill=white,size=2](x1,w1,01,xx1,z_{3},z_{2},z_{1})
 \tkzDrawPoints[color=black,fill=black,size=2](OO,OOO)

  \draw (5.45, -1.04) node {$w_1$};
  \draw (6.55, -1.04) node {$w_2$};

  \draw (0.85, -0.33) node {$w_1$};
  \draw (2.1, -1.15) node {$w_2$};
  \draw (2.95, -0.75) node {$w_3$};
\end{tikzpicture}
\vspace{-2cm}
\caption{ (a) $N=3$ ideal tree, \; (b)  $N=3$ non-ideal tree. The outer triangles are depicted in dashed lines, different colours correspond to different weights. }
\label{Steiner_twb_pic}
\end{figure}

In this section, the lengths of $N = 3$ ideal and non-ideal Steiner trees and $N=4$ ideal Steiner tree are found for arbitrary weights. On the other hand,  $N=4, 5$ non-ideal trees are considered in the superlight approximation. We generalize this approach to higher $N$ and consider a particular example of the $N$-point non-ideal Steiner tree.
\vspace{-1.5mm}
\paragraph{N=3 trees.}
\label{subsubsec:ex}
Let us consider $ N = 3 $ ideal Steiner tree with three boundary endpoints $w_i$ and outer edges of lengths $X_i\;, i = 1,2,3 $ (see (a) Fig. \bref{Steiner_twb_pic}).

  Since the Steiner tree splits the outer triangle into three triangles with two vertices on the boundary we apply \eqref{cos_th2} to each of them and find
 \be
 \label{n3s}
 \ba{c}
 \vspace{1.9mm}
     \dps a_{12} = X_1 + X_2 + 2 \log \sin \frac{\gamma_{12}}{2} \;,\\
\vspace{1.9mm}
    \dps    a_{23} = X_3 +X_2 +  2 \log \sin \frac{\gamma_{23}}{2} \;, \\
    \dps     a_{13} = X_3 + X_1 +  2 \log \sin \frac{\gamma_{13}}{2} \;,
        \ea
 \ee where $a_{ij}$  and $\gamma_{ij}$  are given by \eqref{r1} and \eqref{cooos}. Solving this system of linear equations we find the weighted length defined by \eqref{minimum} as
 \be
 \label{id3}
 L^{(3)}_{\mathbb{D}} (w_i| \epsilon_i) = (\epsilon_1 + \epsilon_2 - \epsilon_3)\log \sin w_{21} + (\epsilon_1 + \epsilon_3 - \epsilon_2)\log \sin w_{31}  +
 (\epsilon_3 + \epsilon_2 - \epsilon_1)\log \sin w_{32} + C  \;,
 \ee
 where
\be
C =  2 \left(\log \sin \frac{\gamma_{12}}{2}  + \log \sin \frac{\gamma_{23}}{2} +  \log \sin \frac{\gamma_{13}}{2} \right)  \;.
\ee

\begin{figure}[H]
\centering
\begin{tikzpicture}[scale=2.5]

  \tkzDefPoint(0,0){O}
  \tkzDefPoint(1,0){A}
  \tkzDrawCircle[line width  = 0.3](O,A)

  \tkzDefPoint(-1,0.001){w_1}
  \tkzDefPoint(-0.661,-0.75){w_2}
  \tkzDefPoint(0.835,-0.55){w_4}
  \tkzDefPoint(0.52,-0.85){w_3}

  \tkzDefPoint(-0.45,-0.25){F1}
  \tkzDefPoint(0.54,-0.53){F2}

  \tkzDefPoint(-1.3,-0.19){c2}
  \tkzDefPoint(-1, -0.73){c1}
  \tkzDefPoint(0.735, -0.699){c4}
  \tkzDefPoint(0.67, -0.8){c3}

   \tkzDefPoint(-0.2, -1.25){cb}
\tkzDefPoint(-0.11,-1.25){d1}
\tkzDefPoint(-0.78,-2.6){d2}
\tkzDefPoint(-1.11,-0.5){d3}
\tkzDefPoint(0.8,-0.82){d4}

\tkzDrawArc[rotate,dashed,color=black,line width  = 0.3](d3,w_1)(-103)
\tkzDrawArc[rotate,dashed,color=black,line width  = 0.3](d2,w_1)(-43)
\tkzDrawArc[rotate,dashed,color=black,line width  = 0.3](d1,w_2)(-105)
\tkzDrawArc[rotate,dashed,color=black,line width  = 0.3](d4,w_3)(-95)

 \tkzDrawArc[rotate,color=blue,line width  = 0.9](c2,F1)(-37)
 \tkzDrawArc[rotate,color=brown,line width  = 0.9](c1,F1)(49)
 \tkzDrawArc[rotate,color=green,line width  = 0.9](c4,F2)(77)
 \tkzDrawArc[rotate,color=pink,line width  = 0.9](c3,F2)(-61)

 \tkzDrawArc[rotate,color=red,thick](cb,F2)(60)
 \tkzDrawPoints[color=black,fill=white,size=2](O,w_1,w_2,w_3,w_4,F1,F2)
 \tkzDrawPoints[color=black,fill=black,size=2](O)

 \draw (-0.7, -0.95) node {$w_2$};
  \draw (-1.2, 0) node {$w_1$};
  \draw (0.52, -1.03) node {$w_3$};
  \draw (1.09, -0.6) node {$w_4$};

   \tkzDefPoint(3,0){OO}
  \tkzDefPoint(4,0){AA}
  \tkzDrawCircle[line width  = 0.3](OO,AA)

  \tkzDefPoint(2,0.001){ww_1}
  \tkzDefPoint(2.339,-0.75){ww_2}
  \tkzDefPoint(3.835,-0.55){ww_4}
  \tkzDefPoint(3.52,-0.85){ww_3}

  \tkzDefPoint(2.55,-0.25){FF1}
  \tkzDefPoint(3.54,-0.53){FF2}

  \tkzDefPoint(1.7,-0.19){cc2}
  \tkzDefPoint(2, -0.73){cc1}
  \tkzDefPoint(3.735, -0.699){cc4}
  \tkzDefPoint(3.67, -0.8){cc3}

   \tkzDefPoint(2.8, -1.25){ccb}
\tkzDefPoint(3.1,-1.48){dd1}
\tkzDefPoint(2.2,-1.95){dd2}
\tkzDefPoint(1.89,-0.5){dd3}
\tkzDefPoint(3.8,-0.82){dd4}

\tkzDefPoint(2.89,-1.25){ad1}
\tkzDefPoint(2.22,-2.6){ad2}

\tkzDrawArc[rotate,color=black,line width  = 0.3](dd3,ww_1)(-103)
\tkzDrawArc[rotate,color=black,line width  = 0.3](dd2,ww_1)(-49)
\tkzDrawArc[rotate,color=black,line width  = 0.3](dd1,ww_2)(-71)
\tkzDrawArc[rotate,dashed,color=black,line width  = 0.3](dd4,ww_3)(-95)

\tkzDrawArc[rotate,dashed,color=black,line width  = 0.3](ad2,ww_1)(-43)
\tkzDrawArc[rotate,dashed,color=black,line width  = 0.3](ad1,ww_2)(-105)

 \tkzDrawArc[rotate,color=blue,line width  = 0.9](cc2,FF1)(-37)
 \tkzDrawArc[rotate,color=brown,line width  = 0.9](cc1,FF1)(49)
 \tkzDrawArc[rotate,color=green,line width  = 0.9](cc4,FF2)(77)
 \tkzDrawArc[rotate,color=pink,line width  = 0.9](cc3,FF2)(-61)

 \tkzDrawArc[rotate,color=red,thick](ccb,FF2)(60)
 \tkzDrawPoints[color=black,fill=white,size=2](OO,ww_1,ww_2,ww_3, ww_4, FF1,FF2)
 \tkzDrawPoints[color=black,fill=black,size=2](OO)

 \draw (2.3, -0.95) node {$w_2$};
  \draw (1.8, 0) node {$w_1$};
   \draw (3.52, -1.03) node {$w_3$};
  \draw (4.09, -0.6) node {$w_4$};


  \end{tikzpicture}
  \vspace{-3.5cm}
\caption{ (a) $N=4$ ideal tree with five independent weights depicted in different colors and outer tetragon in dashed lines.   (b) The auxiliary triangle (in black lines) dissecting the outer tetragon. }
\label{N4}
\end{figure}
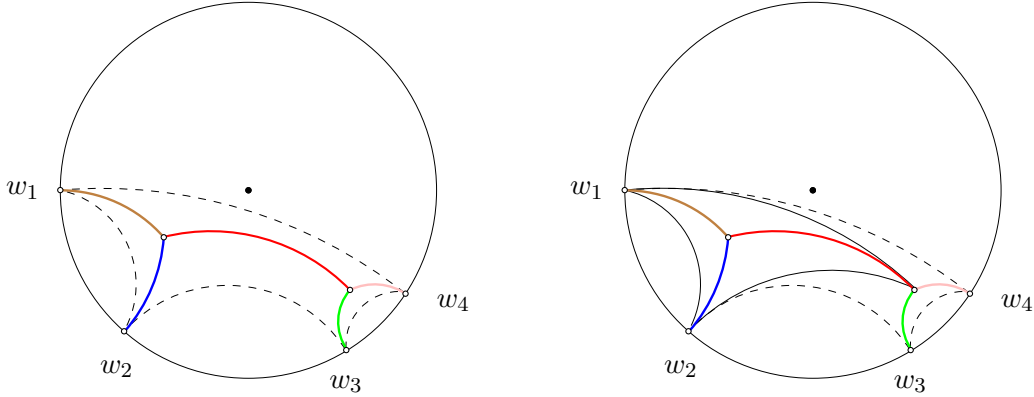

A similar analysis in the case of non-ideal $N=3$ tree is a bit more complicated (see (b) Fig. \bref{Steiner_twb_pic}). Let $Y$ and $Z$ be the lengths of outer edges of weights $\epsilon_{1,2} $ and $X$ be the length of the radial line of weight $\epsilon_3$.  The outer triangle has a vertex in the center of $\mathbb{D}$ and two boundary vertices $w_{1, 2}$.  In this case, the outer triangle is cut by the Steiner tree into two triangles with one boundary vertex  and one triangle with two boundary vertices. Again, using \eqref{cos_th1} and \eqref{cos_th2} we find
\be
\ba{c}
\vspace{2mm}
\dps    2 = \exp[Y] (\cosh X - \sinh X \cos  \gamma_{13} )\;,\\
\vspace{2mm}
\dps     2 = \exp[Z](\cosh X - \sinh X \cos \gamma_{23}) \;,\\
 \dps    a_{12} = Y + Z + 2 \log \sin \frac{\gamma_{12}}{2} \;. \\
\ea
\ee The weighted length of the $N=3$ non-ideal tree  is found to be \footnote{Originally, this length was obtained  in the context of the wordline approach \cite{Hijano:2015rla}. For the analysis in the context of Steiner trees see \cite{Alkalaev:2018nik}.}
\be
\ba{c}
\label{nid3}
\dps L^{(3)}_{\mathbb{D}} (w_{21}|\epsilon_{i}) = \frac{ \epsilon_3}{2} \left[ \text{Arcth}\left[\frac{\cos w_{21}}{\sqrt{1 - \beta^2 \sin^2 w_{12}}}\right] +\gamma \log\sin w_{21}  \right]  \\
\\
\dps - \frac{\epsilon_3 \beta }{2} \log \left(\beta \cos w_{21} + \sqrt{1 - \beta^2 \sin^2 w_{21}}\right) + \tilde C \;,
\ea
\ee where
\be
\gamma = \frac{\epsilon_1 + \epsilon_2}{\epsilon_3}\;, \qquad \beta = \frac{\epsilon_1 - \epsilon_2}{\epsilon_3}\;,
\ee and  $\tilde C$ is given by
\be
\tilde{C} =   \frac{\epsilon_3}{2}\left(\log \frac{\gamma-1}{(\gamma+1)(1-\beta^2)} + \gamma \log \frac{\gamma^2-\beta^2}{(\gamma^2-1)(1-\beta^2)} +\beta \log  \frac{\gamma+\beta}{(1-\beta^2)(\gamma-\beta)}\right)\;.
\ee
\paragraph{Ideal $N=4$ tree.}
\label{subsec:4}
Here we consider an $N=4$ ideal tree with two FT points (see (a) Fig. \bref{N4}). This Steiner tree has four outer edges with weights $\epsilon_i\;, i = 1,...,4$ and one inner edge with weight $\tilde \epsilon$ connecting two FT points.\footnote{The
particular case of the tree with weights $\epsilon_1 = \epsilon_2$ and $\epsilon_3 = \epsilon_4$ was studied in \cite{Alkalaev:2020kxz}.} The  minimum length condition here is encoded by six angles $\alpha_k$ (three at each of the FT points) given by \eqref{cooos}.

Let $A, B, C, D$ denote the regularized lengths of outer edges and $R$ be the length of the inner edge. Consider an auxiliary triangle whose vertices are two boundary endpoints $w_1$ and $w_2$ and the FT point (see (b) Fig. \bref{N4}). Here, $K_1$ and $K_2$ are the regularized lengths of the sides attached to the boundary points $w_1$ and $w_2$ and $\lambda$ and $\lambda'$ are the angles between $R$ and $K_1$ and $K_2$, respectively. Using the relations \eqref{cos_th1}
and \eqref{cos_th2} one finds
\be
\label{lv}
\ba{c}
\exp[K_1] = (\cosh R - \sinh R \cos \alpha_1)\exp[A]\;, \qquad \exp[K_2] = (\cosh R - \sinh R \cos \alpha_2)\exp[B]\;,
\\
\\
\dps A + B + 2 \log \sin \frac{\alpha_3}{2} = a_{12}\;.
\ea
\ee Since $K_{1,2}$ together with the edges $C, D$ cut the outer tetragon into three triangles, one has
\be
\label{conn}
\ba{c}
\exp[K_1 D] (1-\cos(\alpha_4 - \lambda)) = a_{23}\;, \qquad \exp[K_2 C] (1-\cos(\alpha_5 - \lambda')) = a_{14}\;,
\\
\\
\dps \exp[A] = \exp[K_1](\cosh R - \sinh R \cos \lambda)\;, \qquad \exp[B] = \exp[K_2](\cosh R - \sinh R \cos \lambda')\;,
\\
\\
\dps C + D + 2 \log \sin \frac{\alpha_3}{2} = a_{34}\;.
\ea
\ee
Eliminating  $K_{1,2}$ and $\lambda, \lambda'$ from equations \eqref{lv} and \eqref{conn} we obtain
\be
\label{bri}
R = \log \left[ \sqrt{\frac{\gamma_1-1}{\gamma_1+1}}\sqrt{\frac{\gamma_2-1}{\gamma_2+1}} \frac{ \left(1 + 2U -\beta_1 \beta_2+\sqrt{(\beta_1 - \beta_2)^2  + 4U (U +1 - \beta_1 \beta_2)}\right)}{\sqrt{(1 -\beta^2_1)(1 -\beta^2_2} )}\right] \;,
\ee
where
\be
\ba{c}
 \dps U\equiv \exp[\half (a_{23} + a_{14} - a_{34} - a_{12})] = \frac{\sin w_{41} \sin w_{32}}{\sin w_{43} \sin w_{21}}\;, \\
 \\
\dps \gamma_1 = \frac{\epsilon_1 + \epsilon_2}{\tilde \epsilon}\;, \quad \gamma_2 = \frac{\epsilon_3 + \epsilon_4}{\tilde \epsilon} \;, \quad \beta_1 = \frac{\epsilon_1 - \epsilon_2}{\tilde \epsilon} \;,  \quad \beta_2 = \frac{\epsilon_3 - \epsilon_4}{\tilde \epsilon}  \;.
\ea
\ee
The lengths of the outer edges can be found from \eqref{lv}  and \eqref{conn} together with
\be
\dps \exp[K_2 D] (1-\cos(\alpha_4 + \lambda')) = a_{24}\;, \qquad \exp[K_1 C] (1-\cos(\alpha_5 + \lambda)) = a_{13}\;.
\ee
Finally, the weighted length \eqref{minimum} takes the form
\vspace{-2mm}
\be
\label{gen4}
\ba{c}
\dps L^{(4)}_{\mathbb{D}}(w_{i}| \epsilon_{i}, \tilde{\epsilon}) = \tilde \epsilon \left(\gamma_2 \log \sin w_{43} + \gamma_1 \log \sin w_{21} + R \right) -  \\
\\
\dps \frac{\tilde \epsilon (\beta_1 + \beta_2)}{2} \log \left(\frac{2 - \beta^2_1 - \beta^2_2 + 2U (1+ \beta_1 \beta_2) - (\beta_1+ \beta_2) \sqrt{(\beta_1 - \beta_2)^2  + 4U (U +1 - \beta_2 \beta_1)}}{ \sin w_{42} (\sin w_{31})^{-1} (1 + U)}\right) +
\\
\\
\dps \frac{\tilde \epsilon (\beta_2 - \beta_1)}{2} \log \left( \frac{ 2U(\beta_1 \beta_2 -1) - (\beta_1 - \beta_2)^2 +  (\beta_2-  \beta_1) \sqrt{(\beta_1 - \beta_2)^2  + 4U (U +1 - \beta_2 \beta_1)}}{  (\sin w_{41})^{-1} \sin w_{32} \;U }  \right)\;,
\ea
\ee  where we dropped the weight-dependent constants. In the case $\beta_1 = \beta_2 = 0$, which corresponds to equal dimensions $\epsilon_1 = \epsilon_2$ and $\epsilon_3 = \epsilon_4$, the length is given by
\be
\label{N4_simplest}
 L^{(4)}_{\mathbb{D}}(w_{i}| \epsilon_{1}, \epsilon_3, \tilde{\epsilon}) = 2 \epsilon_1 \log \sin w_{43} + 2 \epsilon_3 \log \sin w_{21} + 2 \tilde \epsilon \log (\sqrt{1+ U} + \sqrt{U}) \;.
 \ee

\subsection{$N=4$ and $N=5$ non-ideal trees in the superlight approximation}
\label{subsec:light}
\begin{figure}[H]
\centering
\begin{tikzpicture}[scale=2.0]
 \tkzDefPoint(1.5,0){OO}
  \tkzDefPoint(2.5,0){B}
  \tkzDrawCircle[line width  = 0.3](OO,B)

  \tkzDefPoint(-1.0,0){OOO}
  \tkzDefPoint(0.0,0){C}
  \tkzDrawCircle[line width  = 0.3](OOO,C)

    \tkzDefPoint(4.0,0){O}
  \tkzDefPoint(5.0,0){A}
  \tkzDrawCircle[line width  = 0.3](O,A)


  \tkzDefPoint(-1.945,-0.29){zz_{1}}
  \tkzDefPoint(-1.445,-0.886){zz_{2}}
  \tkzDefPoint(-1.0,-1.0){zz_{3}}

    \tkzDefPoint (-1.825, -0.7){c_c_1}

 \tkzDrawArc[rotate,color=red,line width  = 1.0](c_c_1,zz_{1})(-133)

\draw[color = red, line width  = 1.0] (-1.0, 0) -- (-1.0,-1);

  \tkzDefPoint(0.555,-0.29){zzz_{1}}
  \tkzDefPoint(1.055,-0.886){zzz_{2}}
  \tkzDefPoint(1.5,-1.0){zzz_{3}}

    \tkzDefPoint (0.675, -0.7){cc_c_1}

 \tkzDrawArc[rotate,color=red,line width  = 1.0](cc_c_1,zzz_{1})(-133)

\draw[color = red, line width  = 1.0] (1.5, 0) -- (1.5,-1);

  \tkzDefPoint(3.055,-0.29){k_{1}}
  \tkzDefPoint(3.555,-0.886){k_{2}}
  \tkzDefPoint(4.0,-1.0){k_{3}}
   \tkzDefPoint(4.945,-0.29){k_{4}}
  \tkzDefPoint(4.45,-0.886){k_{5}}

    \tkzDefPoint (3.175, -0.7){k_c_1}
    \tkzDefPoint (4.825, -0.7){k_c_2}

 \tkzDrawArc[rotate,color=red,line width  = 1.0](k_c_1,k_{1})(-133)
 \tkzDrawArc[rotate,color=red,line width  = 1.0](k_c_2,k_{5})(-133)


     \tkzDefPoint (1.09, -0.6){f1}
     \tkzDefPoint (1.5, -1.35){fa}
          \tkzDefPoint (1.5, -0.5){f2}

 \tkzDrawArc[rotate,color=green,line width  = 1.0](fa,f1)(-29)

  \tkzDefPoint (3.59, -0.6){f3}
     \tkzDefPoint (4.0, -1.35){fa1}
    \tkzDefPoint (4.0, -0.5){f4}

\tkzDefPoint (4.41, -0.6){f5}

 \tkzDrawArc[rotate,color=green,line width  = 1.0](fa1,f3)(-58)

 \tkzDrawPoints[color=black,fill=white,size=2](zz_{3},zz_{2},zz_{1}, zzz_{1}, zzz_{2}, zzz_{3}, k_{1}, k_{2},  k_{4}, k_{5}, f1, f2, f3, f5)

 \tkzDrawPoints[color=black,fill=black,size=2](OO,OOO, O)

  \draw (-2.1, -0.33) node {$w_1$};
  \draw (-0.97, -1.13) node {$w_3$};
  \draw (-1.45, -1.01) node {$w_2$};

  \draw (0.4, -0.33) node {$w_1$};
  \draw (1.53, -1.13) node {$w_3$};
  \draw (1.05, -1.01) node {$w_2$};

   \draw (2.9, -0.33) node {$w_1$};
  \draw (3.55, -1.01) node {$w_2$};
  \draw (5.1, -0.33) node {$w^{s}_1$};
  \draw (4.45, -1.01) node {$w^{s}_2$};

\end{tikzpicture}
\caption{ Disconnected $N=4$ tree (a) and $N=4$ non-ideal tree  (b). The green line in (b) carries the superlight weight $\tilde \epsilon_1$, the non-deformed tree is shown in red lines. (c) shows an auxiliary bridge tree associated with the $N=4$ non-ideal tree.}
\label{N=2per}
\end{figure}
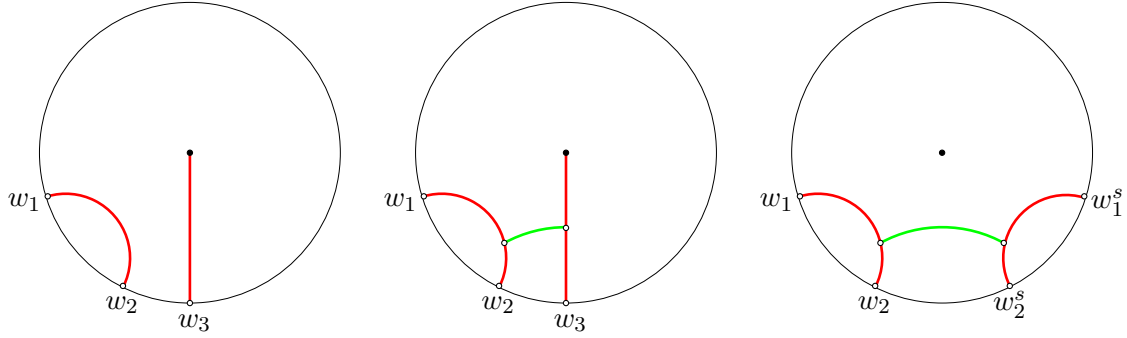

 The lengths of $N \geq 4$ non-ideal Steiner trees with arbitrary weights are unknown. However, the $N=4$ non-ideal tree can be considered as a perturbation of the $N=3$ non-ideal tree with respect to one of outer weights \cite{Alkalaev:2015wia}. In this section we calculate $N=4$ and $N=5$ non-ideal trees in the superlight approximation by perturbing $N=3$ ideal tree and disconnected $N=4$ trees.
 \paragraph{Non-ideal $N=4$ tree from disconnected $N=4$ tree.}
Let us consider a $N=4$ non-ideal tree as a perturbation of a disconnected $N=4$ tree (see (a) and (b) Fig.\bref{N=2per}). The resulting  $N=4$ non-ideal tree has one inner edge with the weight  $\tilde \epsilon_1 \ll \epsilon_{1,3}$ and two pairs of outer edges: the first one with weights $ \epsilon_1 = \epsilon_2 $ is a geodesic connecting $w_1$ and $w_2$ according to \eqref{sl}, and the second one is a radial line with weight $\epsilon_3 = \tilde \epsilon_2$. However, the radial length is a weight-dependent constant so that it can be omitted.

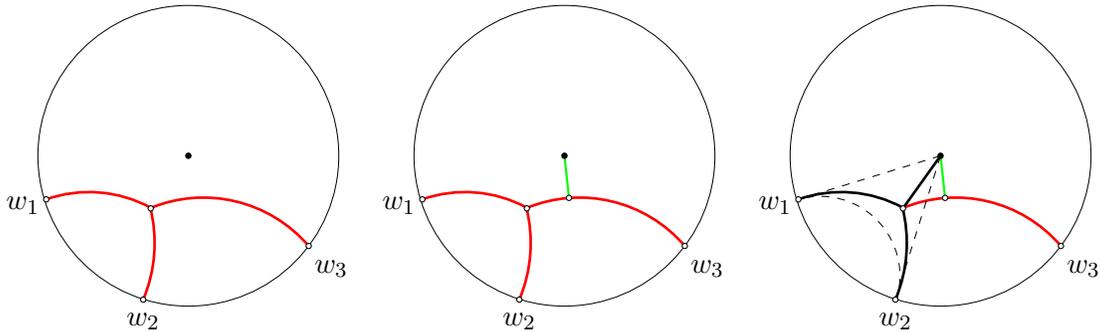
\begin{figure}[H]
\centering
\begin{tikzpicture}[scale=2.]
 \tkzDefPoint(4.5,0){OO}
  \tkzDefPoint(5.5,0){B}
  \tkzDrawCircle[line width  = 0.3](OO,B)

  \tkzDefPoint(2,0){OOO}
  \tkzDefPoint(3,0){C}
  \tkzDrawCircle[line width  = 0.3](OOO,C)

 \tkzDefPoint(7.0,0){O}
  \tkzDefPoint(8.0,0){D}
  \tkzDrawCircle[line width  = 0.3](O,D)

 \tkzDefPoint(2.8,-0.6){z_{3}}
  \tkzDefPoint(1.055,-0.29){z_{1}}
  \tkzDefPoint(1.7,-0.957){z_{2}}
  \tkzDefPoint (1.34, -1.1){c_c_1}
  \tkzDefPoint (0.8, -0.58){c_c_2}
  \tkzDefPoint (2.1,-1.2){c_c_3}

 \tkzDefPoint (1.3, -0){c_o_1}
  \tkzDefPoint (2.537, -0.956){c_o_2}
  \tkzDefPoint (1.62,-2.17){c_o_3}

  \tkzDefPoint(1.75,-0.35){xx1}

  \tkzDrawArc[rotate,color=red,line width  = 1.0](c_c_1,z_{1})(-48)
  \tkzDrawArc[rotate,color=red,line width  = 1.0](c_c_2,z_{2})(37)
  \tkzDrawArc[rotate,color=red,line width  = 1.0](c_c_3,z_{3})(72)

  \tkzDefPoint(5.3,-0.6){zr_{3}}
  \tkzDefPoint(3.555,-0.29){zr_{1}}
  \tkzDefPoint(4.2,-0.957){zr_{2}}
  \tkzDefPoint (3.84, -1.1){cr_c_1}
  \tkzDefPoint (3.3, -0.58){cr_c_2}
  \tkzDefPoint (4.6,-1.2){cr_c_3}

  \tkzDefPoint(4.25,-0.35){xx1r}
  \tkzDefPoint(4.53,-0.28){t};

  \tkzDrawArc[rotate,color=red,line width  = 1.0](cr_c_1,zr_{1})(-48)
  \tkzDrawArc[rotate,color=red,line width  = 1.0](cr_c_2,zr_{2})(37)
  \tkzDrawArc[rotate,color=red,line width  = 1.0](cr_c_3,zr_{3})(72)

\draw[color = green, line width  = 0.8] (4.5, 0) -- (4.53,-0.28);

 \tkzDefPoint(7.8,-0.6){rz_{3}}
  \tkzDefPoint(6.055,-0.29){rz_{1}}
  \tkzDefPoint(6.7,-0.957){rz_{2}}
  \tkzDefPoint (6.34, -1.1){rc_c_1}
  \tkzDefPoint (5.8, -0.58){rc_c_2}
  \tkzDefPoint (7.1,-1.2){rc_c_3}
  \tkzDefPoint (6.2,-0.8){d}

  \tkzDefPoint(6.75,-0.35){xx2r}

  \tkzDrawArc[rotate,color=black,line width  = 1.0](rc_c_1,rz_{1})(-48)
  \tkzDrawArc[rotate,color=black,line width  = 1.0](rc_c_2,rz_{2})(37)
  \tkzDrawArc[rotate,color=red,line width  = 1.0](rc_c_3,rz_{3})(72)

    \tkzDrawArc[rotate,dashed,color=black,line width  = 0.3](d,rz_{1})(-125)

\draw[color = green, line width  = 0.8] (7.0, 0) -- (7.03,-0.28);

\draw[color = black, dashed, line width  = 0.3] (7, 0) -- (6.055,-0.29);
\draw[color = black, dashed, line width  = 0.3] (7, 0) -- (6.7,-0.957);
\draw[color = black, line width  = 1.0] (7, 0) --  (6.75,-0.35);

\tkzDefPoint (7.03,-0.28){xf2}

 \tkzDrawPoints[color=black,fill=white,size=2](xx1, xx1r, xx2r, t, z_{3},z_{2},z_{1}, zr_{3},zr_{2},zr_{1}, xf2, rz_{1}, rz_{2},rz_{3})
 \tkzDrawPoints[color=black,fill=black,size=2](OO,OOO,O)

     \draw (3.4, -0.33) node {$w_1$};
  \draw (4.2, -1.1) node {$w_2$};
  \draw (5.45, -0.75) node {$w_3$};

  \draw (0.9, -0.33) node {$w_1$};
  \draw (1.7, -1.1) node {$w_2$};
  \draw (2.95, -0.75) node {$w_3$};

  \draw (5.9, -0.33) node {$w_1$};
  \draw (6.7, -1.1) node {$w_2$};
  \draw (7.95, -0.75) node {$w_3$};

\end{tikzpicture}
\vspace{-2cm}
\caption{ (a) $N=3$ ideal tree,  (b) $N=4$ non-ideal tree with the superlight weight $\tilde{\epsilon}_2$. The green line represents a perpendicular to the inner edge. (c) An auxiliary Steiner tree shown in black lines, the outer triangle shown in dashed lines.}
\label{N=3per}
\end{figure}

The $N=4$ non-ideal tree without the radial line can be obtained by cutting an auxiliary $N=4$ ideal tree as shown on (c) of Fig.\bref{N=2per}. Such an auxiliary tree has four outer edges with weights $\epsilon_1$ and the outer vertices of the tree are located at points $ (w_1, w_2, w^s_2, w^s_1)$, where $w^{s}_2 = w_1 + 2w_3 - w_2$ and $w^{s}_1 = 2 w_3 + w_1$ are identified by reflecting endpoints $w_1$ and $w_2$ relative to the radius connecting the center of $\mathbb{D}$ and the endpoint $w_3$. Using \eqref{N4_simplest} we find that the length of the $N=4$ non-ideal tree takes the form
\be
 \label{51}
L^{(4)}_{\mathbb{D}} (w_i| \epsilon_1, \tilde \epsilon_1) = 2 \epsilon_1 \log \sin w_{21} + \tilde \epsilon_1  \log (\sqrt{1+ \tilde U} + \sqrt{\tilde U})\;, \quad \tilde U = \frac{\sin (w_3 - w_2) \sin w_3 }{\dps \sin^2 \frac{w_2 - w_1}{2}}\;.
\ee
\paragraph{Non-ideal $N=4$ tree from ideal $N=3$ tree.}
\label{subsec:3N3}
Another example of a $N=4$ non-ideal tree is obtained by adding an outer edge with superlight weight $\tilde \epsilon_2$  to the $N=3$ ideal tree (see (b) Fig.\bref{N=3per}). According to \eqref{lv} the outer edge (denoted by $K$) is the perpendicular to the third edge of the $N=3$ ideal tree.

Let us consider an auxiliary triangle with two boundary vertices and a third vertex in the center of $\mathbb{D}$ (see (c) Fig.\bref{N=3per}). An auxiliary Steiner tree of the triangle consists of edges $X_1, X_2$ of the $N=3$ ideal tree and the edge $A$ stretched to the center of $\mathbb{D}$. To simplify the calculations here we assume $\epsilon_1 = \epsilon_2$. Using the trigonometric relations \eqref{cos_th} and \eqref{cos_th1} we find \be
\ba{c}\label{ot}
\exp[X_1]( \cosh A - \sinh A \cos ( \gamma_{13} + \alpha ))  = 2\;, \quad \exp[X_2]( \cosh A - \sinh A \cos ( \gamma_{13} - \alpha ))  = 2\;, \\
\\
 \sinh K \sin \alpha =  \sinh A \;,
\ea
\ee
where $\alpha$ is the angle between edges $X_3$ and $A$. Solving equations \eqref{ot} in the variable $K$ we obtain
\be
\dps \sinh K = \frac{\sin \dps \frac{2 w_3 - w_2 - w_1}{2}}{\sin \dps \frac{w_3 - w_1}{2} \sin \dps \frac{w_3 - w_2}{2}} \;.
\ee
Then, the length of the non-ideal $N=4$ tree takes the form
\be
\label{n33}
 L^{(4)}_{\mathbb{D}} (w_i| \epsilon_{1}, \epsilon_3, \tilde \epsilon_2) =  L^{(3)}_{\mathbb{D}} (w_i|  \epsilon_{1}, \epsilon_1, \epsilon_3) + \tilde \epsilon_2 \; \text{Arcsinh}\; \frac{\sin \dps \frac{2 w_3 - w_2 - w_1}{2}}{\sin \dps \frac{w_3 - w_1}{2} \sin \dps \frac{w_3 - w_2}{2}} \;,
 \ee  where $ L^{(3)}_{\mathbb{D}} (w_i| \epsilon_1, \epsilon_1, \epsilon_3)$  is given by \eqref{id3}.

\begin{figure}[H]
\centering
\begin{tikzpicture}[scale=2.0]

    \tkzDefPoint(7,0){O}
  \tkzDefPoint(8,0){A}
  \tkzDrawCircle[line width  = 0.3](O,A)

  \tkzDefPoint(6.055,-0.29){l_{1}}
  \tkzDefPoint(6.555,-0.886){l_{2}}
   \tkzDefPoint(7.945,-0.29){l_{4}}
  \tkzDefPoint(7.445,-0.886){l_{5}}

    \tkzDefPoint (6.175, -0.7){l_c_1}
    \tkzDefPoint (7.825, -0.7){l_c_2}

 \tkzDrawArc[rotate,color=red,line width  = 1.0](l_c_1,l_{1})(-133)
 \tkzDrawArc[rotate,color=red,line width  = 1.0](l_c_2,l_{5})(-133)

\tkzDefPoint (6.59, -0.6){f3}
     \tkzDefPoint (7.0, -1.35){fa1}
    \tkzDefPoint (7.5, -0.43){F4}

\tkzDefPoint (7.41, -0.6){f5}

 \tkzDrawArc[rotate,color=green,line width  = 1.0](fa1,f3)(-58)

  \draw[color = green, line width  = 1.0] (7, 0) --  (7.5, -0.43);

 \tkzDrawPoints[color=black,fill=white,size=2](l_{1}, l_{2},   l_{4}, l_{5}, f3, F4, f5)

 \tkzDrawPoints[color=black,fill=black,size=2]( O)

   \draw (5.9, -0.33) node {$w_1$};
  \draw (6.55, -1.01) node {$w_2$};
  \draw (8.1, -0.33) node {$w_4$};
  \draw (7.45, -1.01) node {$w_3$};

\end{tikzpicture}
\caption{ $N=5$ non-ideal tree. The unperturbed $N=4$ tree is shown in red. The green lines represent the inner edge of the tree and the radial line with superlight weights $\tilde \epsilon_1$ and $\tilde \epsilon_3$, respectively. }
\label{N=5}
\end{figure}
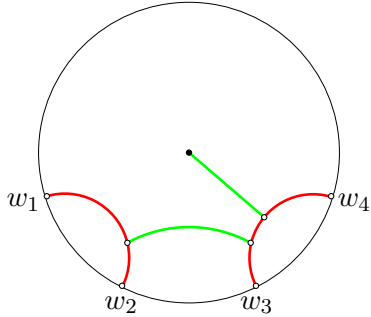

\paragraph{Non-ideal $N=5$ tree from $N=4$ disconnected tree.}
Here, we consider a $N=5$ non-ideal tree with two superlight weights $\tilde \epsilon_{1}, \tilde \epsilon_{3}$, see Fig.\bref{N=5}. The unperturbed $N=4$ disconnected tree is given by two geodesics with weights $\epsilon_1, \epsilon_3$ connecting pairs $w_1, w_2$ and $w_3, w_4$, respectively.
In the superlight approximation the length of the tree is given by the sum of the length of the $N=4$ ideal tree  and the length of the radial line. The length of the radial line given by the first term in \eqref{nid3} under the condition $\tilde{\epsilon}_3 \ll \epsilon_{1,3}$ is equal to
\be
\label{r}
 L^{(r)}_{\mathbb{D}} (w_{43}) = \tilde{\epsilon}_3 \log \cot \dps \frac{w_{43}}{2}\;.
\ee The length of the bridge line with weight $\tilde{\epsilon}_1 \ll \epsilon_{1,3}$ stretched between the geodesics is given by the last terms in formula \eqref{N4_simplest} as
\be
\label{br}
L^{(b)}_{\mathbb{D}} (w_{i})  = \tilde \epsilon_1  \log (\sqrt{1+ U} + \sqrt{ U}) \qquad U = \frac{\sin w_{41} \sin w_{32}}{\sin w_{43} \sin w_{21}}\;.
\ee In this case, the lengths \eqref{r} and \eqref{br} are determined only by coordinates $w_i$ and do not depend on the structure of the unperturbed tree, i.e.weights $\epsilon_{1,3}$.  Finally, the weighted length of the $N=5$ non-ideal tree takes the form
\be
\label{5n}
L^{(5)}_{\mathbb{D}} (w_{i}| \epsilon_1,\epsilon_3, \tilde \epsilon_1, \tilde \epsilon_3) = 2 \epsilon_1 \log \sin w_{21} + 2 \epsilon_3 \log \sin w_{43} + \tilde \epsilon_1  \log (\sqrt{1+ U} + \sqrt{ U}) + \tilde \epsilon_3 \log \cot \dps \frac{w_{43}}{2} \;.
\ee
\paragraph{Multi-point trees.}
\label{subsec:multi}

\begin{figure}[H]
\centering
\begin{tikzpicture}[scale=3.]
  \tkzDefPoint(0,0){O}
  \tkzDefPoint(1,0){A}
  \tkzDrawCircle[line width  = 0.3](O,A)

  \tkzDefPoint(-0.995,-0.1){w_0}
  \tkzDefPoint(-0.979, 0.2){w_m}

  \tkzDefPoint(-0.953,-0.3){w_1}
  \tkzDefPoint(-0.7, -0.714){w_2}

   \tkzDefPoint(-0.527,-0.85){w_3}
   \tkzDefPoint(0.3,-0.953){w_4}

   \tkzDefPoint(0.953,-0.3){w_f}
   \tkzDefPoint(0.8, -0.6){w_pf}

    \tkzDefPoint(-0.88,-0.54){a1}
    \tkzDefPoint(-0.14,-1.1){a2}
    \tkzDefPoint(0.94, -0.48){a3}
    \tkzDefPoint(-1.01, 0.05){a4}

     \tkzDefPoint(-0.99, -0.235){af1}
      \tkzDefPoint(-0.646, -0.83){af2}

       \tkzDefPoint(0.3, -0.835){af3}
      \tkzDefPoint(0.9, -0.8){af4}

\tkzDefPoint(-0.9,-0.054){f_1}
\tkzDefPoint(-0.8,-0.299){f_2}
\tkzDefPoint(-0.634,-0.5){f_3}
\tkzDefPoint(-0.35,-0.69){f_4}
\tkzDefPoint(0.2,-0.79){f_5}
\tkzDefPoint(0.756,-0.5){f_6}
\tkzDefPoint(0.777,-0.39){f_7}




  \tkzDrawArc[rotate,color=red,line width  = 1.0](a1,w_1)(-150)
  \tkzDrawArc[rotate,color=red,line width  = 1.0](a2,w_3)(-130)
  \tkzDrawArc[rotate,color=red,line width  = 1.0](a3,w_pf)(-138)
   \tkzDrawArc[rotate,color=red,line width  = 1.0](a4,w_0)(158)

   \tkzDrawArc[rotate,color=green,line width  = 1.0](af1,f_1)(-86)
   \tkzDrawArc[rotate,color=green,line width  = 1.0](af2,f_3)(-64)

    \tkzDrawArc[rotate,color=green,line width  = 1.0](af3,f_5)(-36)
   \tkzDrawArc[rotate,color=green,line width  = 1.0](af4,f_6)(14)

 \draw[color = green, line width  = 1.0] (0, 0) --  (0.777,-0.39);

 \draw (-1.05, 0.2) node {\scriptsize$w_{_{1}}$};
 \draw (-1.07, -0.13) node {\scriptsize$w_{_{2}}$};

 \draw (-1.05, -0.33) node {\scriptsize$w_{_{3}}$};
 \draw (-0.72, -0.83) node {\scriptsize$w_{_{4}}$};

 \draw (-0.5, -1) node {\scriptsize$w_{_{5}}$};
 \draw (0.35, -1.07) node {\scriptsize$w_{_{6}}$};

 \draw (0.55, -0.96) node {.};
 \draw (0.65, -0.9) node {.};
 \draw (0.74, -0.83) node {.};

\draw (1.07, -0.64) node {\scriptsize$w_{_{2M-1}}$};
\draw (1.2, -0.3) node {\scriptsize$w_{_{2M}}$};

  \tkzDrawPoints[color=black,fill=white,size=2](w_0, w_m, w_1,w_2,w_3,w_4, w_pf, w_f, f_1, f_2, f_3, f_4, f_5, f_6, f_7)
  \tkzDrawPoints[color=black,fill=black,size=2](O)

\end{tikzpicture}
\caption{$N=2M +1$ non-ideal Steiner tree in the superlight approximation. Red lines correspond to connecting outer vertices $w_{2i - 1}$ and $w_{2i}$. The inner edges with weights $\tilde \epsilon_j$ and the radial line with the weight $\epsilon_r$ are shown in green. }
\label{def_dis}
\end{figure}
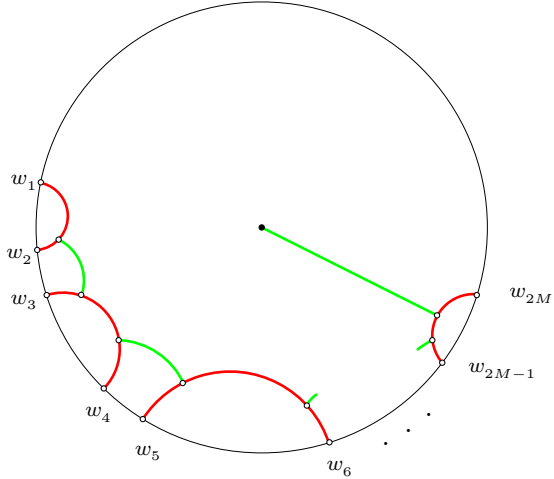

The superlight approximation allows one to calculate the length of a multi-point non-ideal Steiner tree with $N = 2M +1,  \; M = 3, 4, 5,...$ outer vertices. The tree is a perturbation of a disconnected $N = 2M$ Steiner tree consisting of $M$ geodesics with weights $\epsilon_i,  \;i = 1,...,N $ which connect the points $w_{2i-1}, w_{2i}$. The inner bridge lines with superlight weights $\tilde \epsilon_j,\; j = 1,...,N-1$ are connected to the geodesics at FT points and the last outer edge connected to the center of $\mathbb{D}$ carries the weight $\epsilon_r$ (see Fig. \bref{def_dis}).

Since the lengths of the radial and bridge lines for the tree are given by \eqref{r} and \eqref{br}, then the weighted length of the $N = 2M + 1$ non-ideal tree takes the form

\be
\label{2M}
L^{(2M + 1)}_{\mathbb{D}} (w_{i} | \epsilon, \tilde \epsilon) = L^{(2M)}_{\mathbb{D}}  (w_{i} | \epsilon) + 2 \sum^{M-1}_{i=1} \tilde \epsilon_j \log \left(\sqrt{U_{2i +1} + 1 } + \sqrt{U_{2i +1} } \right) + \epsilon_r \log \cot \dps \frac{w_{_{2M-1, 2M}}}{2} \;, \ee
where
\be
\dps L^{(2M)}_{\mathbb{D}}  (w_{i} | \epsilon)  = 2 \sum^{M}_{i=1} \epsilon_i \log \sin w_{2i -1 , 2i} \;, \qquad U_{2 i - 1} = \frac{\sin w_{2i+ 1, 2i} \; \sin  w_{2i+2, 2i - 1}}{\sin  w_{2i, 2i - 1}\; \sin  w_{2i+2, 2i + 1} } \;.
\ee
Note that this analysis can be generalized  to other cases of non-ideal Steiner trees in the superlight approximation. For example, one can consider a disconnected $M=3N$ tree consisting of $N$ ideal Steiner trees  with three boundary endpoints as an unperturbed tree. However, the example is more complicated from a computational point of view.

\section{Large-$c$ conformal blocks}
 \label{sec:semi-CFT}

Here, we discuss the $n$-point large-$c$ conformal blocks with heavy operators in the heavy-light approximation. To this end, we use the monodromy method \cite{Hartman:2013mia, Harlow:2011ny, Fitzpatrick:2014vua, Hijano:2015qja, Banerjee:2016qca, Alkalaev:2016rjl} to demonstrate the holographic correspondence relation \eqref{fS} for particular examples of $5$-point and  $6$-point blocks with superlight operators. This analysis is generalized to the $(2M+2)$-point conformal block.

\subsection{Large-$c$ conformal blocks and monodromy method}
\label{subsec:def}

Consider primary operators  $\cO_i(z_i, \bar z_i),\; i =1,...,n$  at fixed points $(z, \bar{z}) = \{(z_1, \bar z_1), ..., (z_n, \bar z_n)\}$. Let  $F_n(z|\Delta_i, \tilde \Delta_p, c)$ be the corresponding  holomorphic conformal block which depends on conformal dimensions  $\Delta_i$ and exchange dimensions $\tilde \Delta_p \;, p = 1,..., n-3$ and the central charge $c$ \cite{Belavin:1984vu}. Assuming that in the limit $c \rightarrow \infty$  dimensions $\Delta$ and $\tilde{\Delta}$ are proportional to the central charge one can check perturbatively up to a sufficiently high order that the conformal block takes the exponential form
\cite{Zamolodchikov:1987ie} \footnote{Conformal blocks beyond these limits limit are considered in \cite{Fitzpatrick:2015dlt, Kusuki:2018nms}. For recent study of the block exponentiation see \cite{Besken:2019jyw}.}
\be
\label{SEM}
F_n(z| \tilde{\Delta}_p, \Delta_i, c) = \exp\left[\frac{c}{6} f_n(z| \epsilon_i, \tilde{\epsilon}_p)\right] + O\left(\frac{1}{c}\right) \;, \quad \epsilon_{i} \equiv \frac{6 \Delta_i}{c}  \;, \qquad        \tilde{\epsilon}_{p} \equiv \frac{6 \tilde \Delta_p}{c} \;,
\ee where $f(z| \epsilon_i, \tilde{\epsilon}_p)$ is a large-$c$ block, $\epsilon_i, \tilde \epsilon_p$ are classical dimensions which are finite in the large-$c$ limit.
 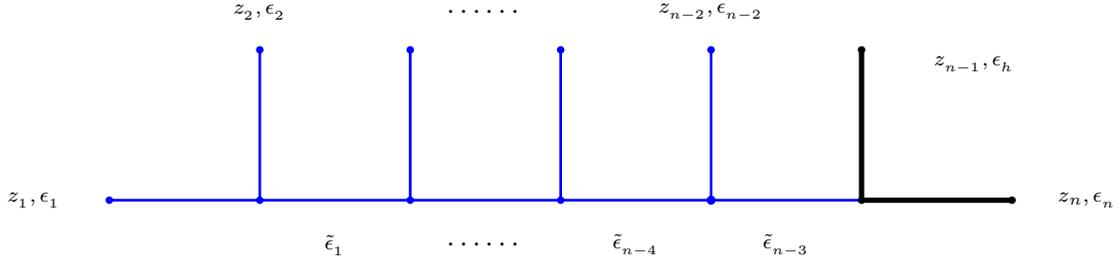
\begin{figure}[H]
\centering
\begin{tikzpicture}[scale=1.0]
\draw [blue, line width=1pt] (30,0) -- (32,0);
\draw [blue, line width=1pt] (32,0) -- (32,2);
\draw [blue, smooth, tension=1.0, line width=1pt] (32,0) -- (34,0);
\draw [blue, line width=1pt] (34,0) -- (34,2);
\draw [blue, smooth, tension=1.0, line width=1pt] (34,0) -- (36,0);
\draw [blue, line width=1pt] (36,0) -- (36,2);
\draw [blue, smooth, tension=1.0, line width=1pt] (36,0) -- (38,0);
\draw [blue, line width=1pt] (38,0) -- (38,2);
\draw [blue, smooth, tension=1.0, line width=1pt] (38,0) -- (40,0);
\draw [line width=2pt] (40,0) -- (40,2);
\draw [line width=2pt] (40,0) -- (42,0);


\draw (29,-0) node {\scriptsize$z_{_{1}}, \epsilon_{_{1}}$};
\draw (32,2.5) node {\scriptsize$z_{_{2}}, \epsilon_{_{2}}$};
\draw (38,2.5) node {\scriptsize$z_{_{n-2}}, \epsilon_{_{n-2}}$};

\draw (35,2.5) node {$\cdots\cdots$};
\draw (43,-0) node {\scriptsize$z{_{n}}, \epsilon_{_{n}}$};
\draw (41.5,1.8) node {\scriptsize$z_{_{n-1}}, \epsilon_{_{h}}$};

\draw (33,-0.6) node {\scriptsize$\tilde\epsilon_{_{1}}$};
\draw (39,-0.6) node {\scriptsize$\tilde\epsilon_{_{n-3}}$};
\draw (37,-0.6) node {\scriptsize$\tilde\epsilon_{_{n-4}}$};
\draw (35,-0.6) node {$\cdots\cdots$};


\fill[blue] (32,0) circle (0.5mm);
\fill[blue] (30,0) circle (0.5mm);
\fill[blue] (32,2) circle (0.5mm);

 \fill[blue] (34,0) circle (0.5mm);
\fill[blue] (34,2) circle (0.5mm);

\fill[blue] (36,0) circle (0.5mm);
\fill[blue] (36,2) circle (0.5mm);

\fill[blue] (38,0) circle (0.6mm);
 \fill[blue] (38,2) circle (0.5mm);

\fill       (40,0) circle (0.5mm);

\fill       (40,2) circle (0.5mm);

\fill       (42,0) circle (0.5mm);
\end{tikzpicture}
\caption{The $n$-point perturbative conformal block with two background operators depicted by bold black lines. }
\label{pbblock}
\end{figure}
 In what follows, we work within the heavy-light approximation \cite{Fitzpatrick:2014vua, Hijano:2015rla, Fitzpatrick:2015zha,  Alkalaev:2016rjl,Alkalaev:2018nik} when two external operators with $\epsilon_{n} = \epsilon_{n-1} = \epsilon_h$ are assumed to be heavier than the other external and exchange operators \footnote{ The case of three or more heavy operators is considered in \cite{Alkalaev:2019zhs, Alkalaev:2020kxz}.} (see Fig.\bref{pbblock})
 \be
 \label{PB}
 \epsilon_i, \tilde \epsilon_p  \ll \epsilon_{h} \;, \quad i = 1,..., n-2 \;, \quad p = 1,..., n-3 \;.
 \ee
\paragraph{Monodromy method and heavy-light approximation.} This method is discussed in details in \cite{Fitzpatrick:2014vua, Hijano:2015rla} for $4$-point conformal blocks and generalized to $n$-point blocks in \cite{Hartman:2013mia, Banerjee:2016qca, Alkalaev:2015wia, Alkalaev:2016rjl}. Below we summarize the main steps.

Let  $\Psi(y|z)$  be an auxiliary $n+1$-point conformal block with one degenerate operator $V_{(1,2)}$ inserted in the point $(y, \bar{y})$ and $n$ primary operators $\cO_{i}$. In the large-$c$ limit the auxiliary block is factored into a product of the form
 \be
 \label{fact}
 \Psi(y|z) \Big|_{c \rightarrow \infty} = \psi(y|z)  \exp\left[\frac{c}{6} f_n(z| \epsilon_i, \tilde{\epsilon}_p)\right] \;,
 \ee where $\psi(y|z)$ is a semiclassical contribution of the operator $V_{(1,2)}$. On the other hand, the auxiliary block satisfies the BPZ equation which is reduced to the Fuchsian-type equation with $n$ singular regular points
 \be
 \ba{c} \label{BPZ}
 \dps \left[\frac{d^2}{dy^2}  + T(y|z)\right]\psi(y|z) = 0\;,
\\
\\
\dps  T(y|z) = \sum_{j=1}^n \frac{\epsilon_j}{(y- z_j)^2} + \frac{c_j}{y- z_j}\;, \qquad c_j = \frac{\partial  f_n(z| \epsilon_i, \tilde{\epsilon}_p)}{\partial z_j}\;,
\ea
\ee where gradients $c_j$ are accessory parameters. In the first order in the heavy-light approximation it leads to the monodromy equations \cite{Alkalaev:2016rjl}
\be
\label{moneq}
I_{+-}^{(n|k)}\,I_{-+}^{(n|k)}+\left(I_{++}^{(n|k)}\right)^2  = -  4\pi^2 \tilde\epsilon^2_{k}\;,
\qquad k = 1,...\,, n-3\,,
\ee
where
$$
\ba{c}
\dps
I^{(n|k)}_{+-}  =\frac{2\pi i}{\alpha}\left[\Big(\alpha \epsilon_1+\sum_{j=2}^{n-2} X_j -\sum_{j=2}^{k+1} (1 - z_j)^{\alpha}(X_j -\epsilon_j \alpha)\right], \qquad \alpha = \sqrt{1 - 4 \epsilon_h}\;,
\ea
$$
\be
\label{Is}
I^{(n|k)}_{-+}  = I^{(n|k)}_{+-}\big|_{\alpha \rightarrow -\alpha}\,,
\qquad\;\;
I^{(n|k)}_{++} = \frac{2\pi i}{\alpha}\sum_{j=k+2}^{n-2} X_j\,, \qquad  X_j = c_j (1-z_j) - \epsilon_j \,.
\ee
These are $n-3$ quadratic equations which can be explicitly solved for lower-point conformal blocks. More specifically, one can find accessory parameters for the $4$-point block \cite{Hijano:2015rla} and $5,6$-identity blocks \cite{Alkalaev:2018nik}, because the system \eqref{moneq} reduces to quadratic equations for each parameter. We will discuss an example of such a $6$-point identity block generalizing the one found in \cite{Alkalaev:2018nik}.\footnote{For a general $5$-point block, the accessory parameters are the roots of fourth-degree equations and the explicit form of such a conformal block is unknown. In the case $n>6$, the solution of the system  \eqref{moneq} can only be written for special factorized blocks. In further one can consider the choice of special values of the classical dimensions and insertion points of operators as a possible way to simplify such a system for multi-point blocks.
}

\paragraph{Holographic correspondence relation.}
The duality between large-$c$ conformal blocks and Steiner trees on the Poincare disk is given by the holographic correspondence relation
\be
\label{fS}
f_n(z_k|\epsilon_k, \tilde\epsilon_p) = -L^{(n-1)}_{\mathbb{D}}(\alpha w_k|\epsilon_k, \tilde\epsilon_p)+ i\sum_{k=1}^{n-2}\epsilon_k w_k\;, \qquad w_k = i \log (1 - z_k) \;,
\ee where $L^{(n-1)}_{\mathbb{D}}(\alpha w_k|\epsilon_k, \tilde\epsilon_p)$ is the weighted length of the Steiner tree corresponding to the $n$-point block with weights $\epsilon_k, \tilde \epsilon_p$ that are equal to the classical dimensions of the block. Note that the length $L^{(n-1)}_{\mathbb{D}}(\alpha w_k|\epsilon_k, \tilde\epsilon_p)$ depends on the rescaled coordinates $\alpha w_k$ due to the fact that $\alpha$ is the angle deficit of the Poincare disk (see Section \bref{sec:st}).

\subsection{Examples of conformal blocks}
\label{subsec:exc}
 In this section we calculate $5$-point and $6$-point large-$c$ conformal blocks dual to the lengths of Steiner trees computed in Section \bref{subsec:ex}. By virtue of \eqref{fS} the relations \eqref{triangle} define the fusion rules for such blocks. We use the following  variables
 \be
 P_{j} = (1-z_j)^{\alpha}\;, \qquad j = 2,..., n-3 \;,
 \ee
and set $w_1 = 0$ in the lengths of Steiner trees due to the condition $P_1 = 1$.

\paragraph{$5$-point non-identity blocks  with superlight operators.}
Here we suppose that one of the exchange operator dimensions $\tilde \epsilon_1$ or $\tilde \epsilon_2$ is superlight: $\tilde \epsilon_{1,2} \ll \epsilon_{1,2,3}$.\footnote{For the analysis of other approximations used to calculate $5$-point large-$c$ block see \cite{Belavin:2017atm, Alkalaev:2015fbw}.} The first example corresponds to $\tilde \epsilon_1 \ll \epsilon_{1,2,3}$. The weighted length of the dual Steiner tree is given by \eqref{51} and from the fusion rules \eqref{triangle} we get $\epsilon_1 = \epsilon_2$ and $\tilde \epsilon_2 = \epsilon_3$. Then, using the holographic correspondence relation \eqref{fS} we find
\be
\ba{c}
\label{b5}
f_5(z|\epsilon_1, \epsilon_3, \tilde \epsilon_1) =   \epsilon_1 (-1 + \alpha)\log P_2^{1/\alpha} - ( \epsilon_3 +  \alpha \tilde \epsilon_1 ) \log P_3^{1/\alpha}  \\
\\
- ( 2 \epsilon_1+ \tilde \epsilon_1 )\log[1 - P_2] + \tilde \epsilon_1 \log [P_2 - P^2_3 - \sqrt{(1 - P^2_3)(P^2_2 - P^2_3)}]  \;.
\ea
\ee
The accessory parameters corresponding to the conformal block \eqref{b5} must satisfy monodromy equations \eqref{moneq}
\be
\label{meq5}
\left(I^{(5|1)}_{++}\right)^2 + I^{(5|1)}_{+-}I^{(5|1)}_{-+} = - 4\pi^{2} \tilde{\epsilon}^2_1 \;, \qquad I^{(5|2)}_{+-}I^{(5|2)}_{-+} =  -  4\pi^{2} \epsilon^2_3 \;.
\ee
A few comments are in order. Since the block \eqref{b5} is linear in $\tilde \epsilon_1$ we consider equations \eqref{meq5} up to the second order in $\tilde \epsilon_1$ inclusively. After direct substitution of the accessory parameters corresponding to the conformal block \eqref{b5}, the first equation is satisfied exactly but the left-hand side of the second equation contains the term which is proportional to $\tilde \epsilon_1^2$.\footnote{This term will be canceled when considering the higher order corrections to the block \eqref{b5}.} In what follow we will refer to a \textit{semi-linear} order (in superlight dimensions) as a situation in which the monodromy equations with superlight dimensions in right-hand sides are satisfied exactly and remaining ones are satisfied in the first order in superlight dimensions. In addition, the analysis of equations \eqref{meq5} only in the first order in $\tilde{\epsilon}_1$ gives one nontrivial equation for two accessory parameters and the second one becomes trivial.

Next we  consider the case  $\tilde \epsilon_2 \ll \epsilon_{1,2,3}$.  The block is dual to the Steiner tree of the length \eqref{n33} so that we assume $\epsilon_1 = \epsilon_2$. The fusion rules \eqref{triangle} require  $\tilde \epsilon_1 = \epsilon_3$. According to \eqref{fS} the confromal block has the form
\be
\label{b55}
\ba{c}
f_5(z|\epsilon_1, \epsilon_3, \tilde \epsilon_2) =   \epsilon_1 (-1 + \alpha) \left(\log P_2^{1/\alpha} + \log P_3^{1/\alpha} \right) - \epsilon_3 \left(
\log [1 - P_3] + \log [P_2 - P_3]\right) \\
\\
\dps - (2 \epsilon_1 - \epsilon_3)  \log [1 - P_2] + \tilde \epsilon_2 \text{Arcsinh} \left[\frac{- i (P_2 - P^2_3)}{(1 - P_3)(P_2 - P_3)}\right]\;,
\ea
\ee
and, substituting the accessory parameters associated with the conformal block into the monodromy equations \eqref{moneq}, we find
\be
\dps \left(I^{(5|1)}_{++}\right)^2 + I^{(5|1)}_{+-}I^{(5|1)}_{-+} = - 4\pi^{2} \epsilon^2_3 \;, \qquad
\dps I^{(5|2)}_{+-}I^{(5|2)}_{-+} = -  4\pi^{2}  \tilde{\epsilon}^2_2  \;.
\ee
As in the previous case the monodromy equations are satisfied in the semi-linear order in $\tilde \epsilon_2$.
\paragraph{$6$-point identity block with light operators. }
Let us consider the $6$-point identity block with $\tilde{\epsilon}_3 = 0$ and denote $\tilde \epsilon = \tilde \epsilon_1$. According to the fusion rules \eqref{triangle} it follows that $\tilde{\epsilon}_2 = \epsilon_4$. The length of the corresponding Steiner tree is given by \eqref{gen4} and according to \eqref{fS} the conformal block takes the form
\be
\ba{c}
\label{gr6}
f_{6}(z|\epsilon_i, \tilde \epsilon) =  \epsilon_2 (\alpha - 1)\log P^{1/\alpha}_2 +  \epsilon_3 (\alpha - 1)\log P^{1/\alpha}_3 +  \epsilon_4 (\alpha - 1)\log P^{1/\alpha}_4
\\
\\
\dps - \frac{\tilde \epsilon}{2} \left(\beta_1 \log P_2 - \beta_2 \log P_3 + \beta_2 \log P_4 - \beta_{+} \log \frac{P_2 - P_4}{1-P_3} - \beta_{-} \log \frac{P_2 -P_3}{1-P_4}\right)
\\
\\
 - \tilde \epsilon \left(\gamma_2 \log [P_3 - P_4] + \gamma_1 \log [1 - P_2] + \log \left[1 + 2 U -\beta _2 \beta_1+\sqrt{\beta_{-}^2 + 4 U^2+4 \left(1-\beta_1 \beta_2\right) U} \right]  \right)
 \\
 \\
 + \dps \frac{\tilde \epsilon \beta_{+} }{2} \log \frac{\left(2 - \beta^2_1 - \beta^2_2 + 2U (1+ \beta_1 \beta_2) - \beta_{+} \sqrt{\beta_{-}^2  + 4U (U +1 - \beta_2 \beta_1)}\right)}{1 + U}  \\
 \\
 \dps + \frac{\tilde \epsilon \beta_{-}}{2} \log \left( U \left(2U(\beta_1 \beta_2 -1) - \beta_{-}^2 +  \beta_{-} \sqrt{\beta_{-}^2  + 4U (U +1 - \beta_2 \beta_1)} \right) \right) \;, \\
 \\
\dps U = \frac{(1- P_4)(P_2 - P_3)}{(1 - P_2)(P_3 - P_4)}\;, \qquad \beta_{\pm} = \beta_{1} \pm \beta_{2} \;.
\ea
\ee
The monodromy equations \eqref{moneq} for the conformal block \eqref{gr6} take the form
\begin{equation}
\label{mon_eq6pt}
\ba{c}
\left(I^{(6|1)}_{++}\right)^2 + I^{(6|1)}_{+-}I^{(6|1)}_{-+} = - 4\pi^{2} \tilde{\epsilon}^2\;, \qquad \left(I^{(6|2)}_{++}\right)^2 + I^{(6|2)}_{+-}I^{(6|2)}_{-+} = - 4\pi^{2} \epsilon^2_4\;,
\\
\\
I^{(6|3)}_{+-}I^{(6|3)}_{-+} = 0 \;,
\ea
\end{equation}
and one can explicitly show that the corresponding accessory parameters satisfy these equations without using superlight approximation.

\paragraph{$6$-point non-identity block with superlight operators. }
Here we discuss the case of non-identity $6$-point block  with $\epsilon_1 = \epsilon_2$ and $\epsilon_3 = \epsilon_4$. The fusion rules \eqref{triangle} constrain the dimensions as $\tilde \epsilon_2 = \epsilon_3$. There are two superlight exchange operators with dimensions $\tilde \epsilon_{1,3} \ll \epsilon_{1,3}$. Using holographic correspondence relation \eqref{fS} and the length of the corresponding Steiner tree \eqref{5n} we find that the conformal block takes the form
\be
\ba{c}
\label{b6}
\dps f_6(z|\epsilon_1, \epsilon_3, \tilde \epsilon_1, \tilde \epsilon_3) =   (-1 + \alpha) \left(\epsilon_1 \log P_2^{1/\alpha} + \epsilon_3 \log P_3^{1/\alpha} + \epsilon_3 \log P_4^{1/\alpha} \right) + \tilde \epsilon_3 \log \frac{\sqrt{P_3} - \sqrt{P_4}}{\sqrt{P_3} + \sqrt{P_4}}   \\
\\
\dps - 2 \epsilon_1 \log [1 - P_2] -  2 \epsilon_3 \log [P_3 - P_4] - 2 \tilde \epsilon_1 \log \left( \sqrt{\frac{(1 -P_3)(P_2 - P_4)}{(1 - P_2)(P_3 - P_4)}} + \sqrt{\frac{(1 - P_4)(P_2 - P_3)}{(1 - P_2)(P_3 - P_4)}}\right)\;.
\ea
\ee
After substituting the corresponding accessory parameters into the monodromy equations
\be
\label{meq6}
\ba{c}
\left(I^{(6|1)}_{++}\right)^2 + I^{(6|1)}_{+-}I^{(5|1)}_{-+} = - 4\pi^{2} \tilde{\epsilon}^2_1  \;, \qquad \left(I^{(6|2)}_{++}\right)^2 + I^{(6|2)}_{+-}I^{(6|2)}_{-+} = - 4\pi^{2} \epsilon^2_3\;, \\
\\
I^{(6|3)}_{+-}I^{(6|3)}_{-+} = -  4\pi^{2} \tilde{\epsilon}^2_3\;,
\ea
\ee
we find that they are satisfied up to the semi-linear order in $\tilde \epsilon_{1,3}$.

\paragraph{$(2M+2)$-point conformal block with superlight operators.}
The foregoing analysis can be generalized to $(2M + 2)$-point conformal block dual to the multi-point Steiner tree \eqref{2M} (see Fig. \bref{def_dis} and \bref{sblock}). It has $M$ exchange superlight operators with weights $\tilde \epsilon_j,\; j = 1, ...., M$. The fusion rules are
\be
\label{fus_M}
\epsilon_{2i - 1} = \epsilon_{2i} = \tilde \epsilon_{2i - 1} \;, \qquad i = 1, ..., M \;.
\ee
 \begin{figure}[H]
\centering
\begin{tikzpicture}[scale=1.0]

\draw [blue, line width=1pt] (28,0) -- (30,0);
\draw [blue, line width=1pt] (30,0) -- (30,2);
\draw [green, line width=1pt] (30,0) -- (32,0);
\draw [blue, line width=1pt] (32,0) -- (32,2);
\draw [blue, smooth, tension=1.0, line width=1pt] (32,0) -- (34,0);
\draw [blue, line width=1pt] (34,0) -- (34,2);
\draw [green, smooth, tension=1.0, line width=1pt] (34,0) -- (36,0);
\draw [blue, line width=1pt] (36,0) -- (36,2);
\draw [blue, smooth, tension=1.0, line width=1pt] (36,0) -- (38,0);
\draw [blue, line width=1pt] (38,0) -- (38,2);
\draw [green, smooth, tension=1.0, line width=1pt] (38,0) -- (40,0);
\draw [line width=2pt] (40,0) -- (40,2);
\draw [line width=2pt] (40,0) -- (42,0);


\draw (27.6,-0.1) node {$\epsilon_{_{1}}$};
\draw (30,2.5) node {$\epsilon_{_{1}}$};
\draw (32,2.5) node {$\epsilon_{_{3}}$};
\draw (38,2.5) node {$\epsilon_{_{n-3}}$};

\draw (35,2.5) node {$\cdots\cdots\cdots\cdots\cdots$};
\draw (42.5,-0) node {$\epsilon_{_{h}}$};
\draw (40.0, 2.5) node {$\epsilon_{_{h}}$};

\draw (31,-0.6) node {$\tilde \epsilon_{_{1}}$};
\draw (33,-0.6) node {$\epsilon_{_{3}}$};
\draw (39,-0.6) node {$\tilde \epsilon_{_{n-3}}$};
\draw (37,-0.6) node {$\epsilon_{_{n-3}}$};
\draw (35,-0.6)  node {$\cdots\cdots\cdots\cdots\cdots$};


\fill[blue] (28,0) circle (0.5mm);
\fill[blue] (30,2) circle (0.5mm);

\fill[blue] (32,0) circle (0.5mm);
\fill[blue] (30,0) circle (0.5mm);
\fill[blue] (32,2) circle (0.5mm);

 \fill[blue] (34,0) circle (0.5mm);
\fill[blue] (34,2) circle (0.5mm);

\fill[blue] (36,0) circle (0.5mm);
\fill[blue] (36,2) circle (0.5mm);

\fill[blue] (38,0) circle (0.6mm);
 \fill[blue] (38,2) circle (0.5mm);

\fill       (40,0) circle (0.5mm);

\fill       (40,2) circle (0.5mm);

\fill       (42,0) circle (0.5mm);
\end{tikzpicture}
\caption{ $(2M+2)$-point large-$c$ conformal block with $M$ superlight operators depicted by green lines. }
\label{sblock}
\end{figure}
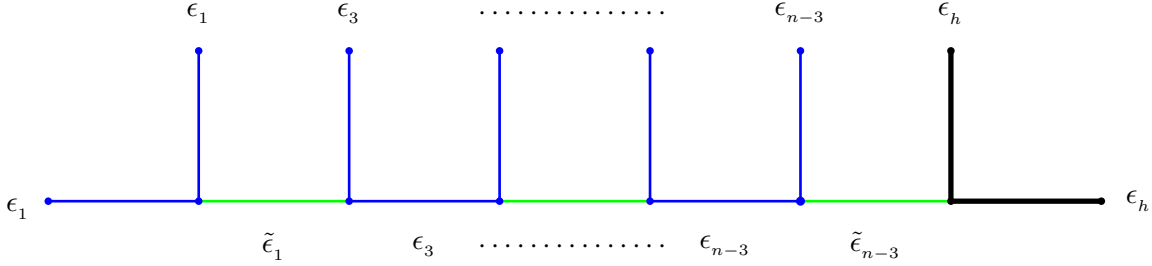 \noindent According to \eqref{fS} and \eqref{2M} the block function takes the form
\be
\label{2d}
\dps f_{2M+2}(z|\epsilon_i, \tilde \epsilon_j) = \sum^{2M}_{i=1} f_{2}(z|\epsilon_i) + \sum^{M - 1 }_{j=1} \tilde{f}_{2}(z|\tilde \epsilon_j) + \tilde \epsilon_{n-3} \log \dps \frac{\sqrt{P_{n-3}} - \sqrt{P_{n-2}}}{\sqrt{P_{n-3}} + \sqrt{P_{n-2}}}\;,
\ee where
\be
\ba{c}
f_{2}(z|\epsilon_i) = (-1 + \alpha) \epsilon_{2i-1} \left( \log P_{2i-1}^{1/\alpha}  + \log P_{2i}^{1/\alpha} \right) - 2 \epsilon_{2i-1} \log [P_{2i - 1} - P_{2i}]\;, \\
\\
\dps \tilde{f}_{2}(z|\tilde \epsilon_j) =  - 2 \tilde \epsilon_{j} \log \left( \sqrt{\frac{(P_{2j-1} -P_{2j+1})(P_{2j} - P_{2j+2})}{(P_{2j-1} - P_{2j})(P_{2j+1} - P_{2j+2})}} + \sqrt{1 + \frac{(P_{2j-1} -P_{2j+1})(P_{2j} - P_{2j+2})}{(P_{2j-1} - P_{2j})(P_{2j+1} - P_{2j+2})}}\right) \;.
\ea
\ee
It can be explicitly shown that this block satisfies the monodromy equations \eqref{moneq} in the semi-linear order in the dimensions of superlight operators $\tilde \epsilon_j$.

\section{Conclusion}
\label{sec:con}
In this paper we explicitly computed the weighted lengths of the $N=4,5,6$ Steiner trees on the Poincare disk and demonstrated that they calculate the dual CFT$_2$ large-$c$ conformal blocks. On the boundary side, the superlight approximation corresponds to superlight operators. Our results along with previously known are shown in the table below. HL and SL denote the heavy-light and the superlight approximations, respectively.

\vspace{0.3cm}
\noindent \begin{tabular}{ || l | l |p{2.7cm} | p{3.6cm}|| }
\hline
$N$ & Steiner tree with $N$ endpoints & $N+1$-point conformal block & ~~~~~approximation\\ \hline
2 & Ideal tree, ref.\cite{Fitzpatrick:2014vua} & ~~~~ref.\cite{Fitzpatrick:2014vua} & ~~~~~~~~~~~HL\\ \hline
3 & Ideal tree, ref.\cite{Alkalaev:2018nik} & ~~~~ref.\cite{Alkalaev:2018nik} & ~~~~~~~~~~~HL\\ \hline
3 & Non-ideal tree, ref.\cite{Hijano:2015rla, Alkalaev:2015wia, Alkalaev:2018nik} & ~~~~ref.\cite{Hijano:2015rla, Alkalaev:2015wia, Alkalaev:2018nik} &  ~~~~~~~~~~~HL\\ \hline
4 & Simplest ideal tree, ref.\cite{Alkalaev:2018nik} & ~~~~ref.\cite{Alkalaev:2018nik}&  ~~~~~~~~~~~HL\\ \hline
4 & General ideal tree, eq.\eqref{gen4} & ~~~~eq.\eqref{gr6} & ~~~~~~~~~~~HL\\ \hline
4 & Non-ideal tree, ref.\cite{Alkalaev:2015lca, Alkalaev:2015wia, Belavin:2017atm} & ~~~~ref.\cite{Alkalaev:2015lca, Alkalaev:2015wia, Belavin:2017atm} & ~~~~~~~~HL+SL \\ \hline
4 & Non-ideal tree, eq.\eqref{51} & ~~~~eq.\eqref{b5} & ~~~~~~~~HL+SL \\ \hline
4 & Non-ideal tree, eq.\eqref{n33}& ~~~~eq.\eqref{b55} & ~~~~~~~~HL+SL \\ \hline
5 & Non-ideal tree, eq.\eqref{5n} & ~~~~eq.\eqref{b6} & ~~~~~~~~HL+SL \\ \hline
N  & Non-ideal tree, ref.\cite{Banerjee:2016qca}&~~~~ref.\cite{Banerjee:2016qca} & ~~~~~~~~HL+SL \\ \hline
N & Various disconnected trees, ref.\cite{Alkalaev:2018nik} & ~~~~ref.\cite{Alkalaev:2018nik} & ~~~~~~~~~~~HL\\ \hline
N = 2M+1  & Non-ideal tree, eq.\eqref{2M}  & ~~~~eq.\eqref{2d} &  ~~~~~~~~HL+SL \\ \hline
\hline
\end{tabular}
\vspace{0.1cm}

 One can analyze $N$-point Steiner trees as deformations of other unperturbed tree configurations.  Also it would be interesting to compute the lengths of Steiner trees in the second and next orders in the superlight approximation.

\vspace{4mm}
\noindent \textbf{Acknowledgements.} I would like to thank K.B. Alkalaev for discussions and the organizers of YRISW 2020 school for the hospitality and productive atmosphere during the visit. The work was supported by RFBR grant No 18-02-01024 and  by the Foundation for the Advancement of Theoretical Physics and Mathematics “BASIS”.

\providecommand{\href}[2]{#2}\begingroup\raggedright\endgroup
\end{document}